\documentclass[3p]{elsarticle}
\usepackage{graphicx}
\usepackage{epsfig}
\usepackage{amssymb}
\usepackage{amsthm}
\usepackage{multirow}
\usepackage[table]{xcolor}
\usepackage{float}
\usepackage{dblfloatfix}
\usepackage{subfigure}
\usepackage{soul}
\usepackage[normalem]{ulem}
\usepackage{url}
\usepackage{hyperref}

\journal{Computer Speech and Language}
\begin{document}

\begin{frontmatter}
\title{rVAD: An Unsupervised Segment-Based Robust \\ Voice Activity Detection Method}
\author{Zheng-Hua Tan$^a$\corref{cor1}}
\address[]{Department of Electronic Systems,
Aalborg University, Denmark \\
 {\small \tt }}
\author{Achintya kr. Sarkar$^{a,b}$}
 \address[]{School of Electronics Engineering, VIT-AP University, India\\
 {\small \tt }}
 \author{Najim Dehak$^c$}
\address[]{Center for Language and Speech Processing, Johns Hopkins University, Baltimore, MD, USA \\
 {\small \tt }}

\cortext[cor1]{Corresponding author. Email address: zt@es.aau.dk. This work was done in part  while the author was visiting Computer Science and Artificial Intelligence Laboratory, Massachusetts Institute of Technology, Cambridge MA, USA.}

\begin{abstract}
This paper presents an unsupervised segment-based method for robust voice activity detection (rVAD). The method consists of two passes of denoising followed by a voice activity detection (VAD) stage. In the first pass, high-energy segments in a speech signal are detected by using \emph{a posteriori} signal-to-noise ratio (SNR) weighted energy difference and if no pitch is detected within a segment, the segment is considered as a high-energy noise segment and set to zero. In the second pass, the speech signal is denoised by a speech enhancement method, for which several methods are explored. Next, neighbouring frames with pitch are grouped together to form pitch segments, and based on speech statistics, the pitch segments are further extended from both ends in order to include both voiced and unvoiced sounds and likely non-speech parts as well. In the end, \emph{a posteriori} SNR weighted energy difference is applied to the extended pitch segments of the denoised speech signal for detecting voice activity. We evaluate the VAD performance of the proposed method using two databases, RATS and Aurora-2, which contain a large variety of noise conditions. The rVAD method is further evaluated, in terms of speaker verification performance, on the RedDots 2016 challenge database 
and its noise-corrupted versions.  Experiment results show that rVAD is compared favourably with a number of existing methods. In addition, we present a modified version of rVAD where computationally intensive pitch extraction is replaced by computationally efficient spectral flatness calculation. The modified version significantly reduces the computational complexity at the cost of moderately inferior VAD performance, which is an advantage when processing a large amount of data and running on low resource devices. The source code of rVAD is made publicly available.  
\end{abstract}


\end{frontmatter}

\section{Introduction}
\label{sec:intro}
Voice activity detection (VAD), also called speech activity detection (SAD), is widely used in real-world speech systems for improving robustness against additive noises or discarding the non-speech part of a signal to reduce the computational cost of downstream processing \cite{price2018low}. It attempts to detect the presence or absence of speech in a segment of an acoustic signal. The detected non-speech segments can subsequently be abandoned to improve the overall performance of these systems. For instance, cutting out noise only segments can reduce the error rates of speech recognition and speaker recognition systems \cite{kenny2014, zt2010, Ramirez2004a}. 
  
VAD methods can be broadly categorized into supervised and unsupervised methods. Supervised methods formulate VAD as a classical classification problem and solve it either by training a classifier directly \cite{zhang2013deep}, or by training statistical models for speech and non-speech separately and then making VAD decisions based on a log-likelihood ratio test (LRT) \cite{ferrernoise2013}. These methods require labelled speech data and their performances are highly dependent on the quality of the labelled data, in particular how well training  and test data match each other. Supervised methods are often able to outperform unsupervised methods under matched conditions, but they can potentially break down under mismatched conditions. In \cite{Ferrer2016}, for example, a deep learning method, experimented on the robust automatic transcription of speech (RATS) database \cite{Walker2012}, demonstrates a very good VAD performance for seen environments, but the gap between unseen and seen environments is very significant, by almost an order of magnitude difference in terms of detection cost function (DCF), but what matters most is the performance on unseen environments. 
Unsupervised VAD methods include metrics methods and model based ones. Metrics methods rely on the continuous observation of a specific metric, such as energy and zero-crossing rate, followed by a simple threshold-based decision stage \cite{petsatodis2011convex}. On the other hand, model based methods build separate models for speech and non-speech, followed by an LRT test together with threshold-based decision for statistical models \cite{petsatodis2011multi} (similarly to the settings of supervised methods but without using labelled data) or a distance comparison for non-statistical models \cite{VQVAD} to classify a segment as speech or non-speech. Semi-supervised VAD has also been studied, e.g., semi-supervised Gaussian mixture models (SSGMM) based VAD for speaker verification in \cite{sholokhov2018semi}.
    
VAD is a binary classification problem involving both feature extraction and classification.  
Various speech features can be found in literature such as energy, zero-crossing rate, harmonicity \cite{chuangsuwanich2011robust}, perceptual spectral flux \cite{Sadjadi2013}, Mel-frequency cepstral coefficient (MFCC) \cite{VQVAD}, power-normalized cepstral coefficients (PNCCs) \cite{Kinnunen+2016}, entropy \cite{Renevey2001a},  Mel-filter bank (MFB) outputs \cite{Vlaj2005a} and \emph{a posteriori} signal-to-noise ratio (SNR) weighted energy distance \cite{zt2010, tan2008posteriori}. For the purpose of modelling and classification, popular techniques include Gaussian models  \cite{Sohn1999a}, Gaussian mixture models (GMM) \cite{kenny2014,Sadjadi2013,Bonastre2004}, super-Gaussian models and their convex combination \cite{petsatodis2011convex}, i-vector \cite{Kinnunen+2016,Khoury+2016}, decision trees \cite{Hu2012b},  support vector machines \cite{Dong2002a}, and neural network models (including deep models) \cite{zhang2013deep, Ferrer2016, Tao2017}.

Although much progress has been made in the area of VAD, developing VAD methods that are accurate in both clean and noisy environments and are able to generalize well under unseen environments is still an unsolved problem. For example, the ETSI advanced front-end (AFE) VAD \cite{ETSI2007}, a commonly referred VAD method, performs very well in noisy environments, but poorly in noise-free conditions, and it is primarily suitable for dealing with stationary noise. 
The MFB output based VAD \cite{Vlaj2005a} and the long-term signal variability (LTSV) VAD \cite{Ghosh2011} are highly accurate for clean speech, but their performances in noisy environments are worse than that of the AFE VAD.  A supervised VAD method based on a non-linear spectrum modification (NLSM) and GMM \cite{Vlaj2012} is shown to be able to outperform both MFB and LTSV algorithms. In \cite{zt2010} a low-complexity VAD method based on \emph{a posteriori} SNR weighted energy difference also shows a performance superior to several methods including the MFB VAD and the AFE VAD. These referred methods are all significantly superior to the G.729 \cite{ITU1996a} and G.723.1 \cite{ITU1996b} VAD algorithms included in their corresponding Voice-over-IP standards. The comparisons cited in the paragraph are all conducted on the Aurora 2 database \cite{Hirsch2000}.  


This paper focuses on developing a VAD method that is unsupervised and able to generalize well to real-world data, irrespective of whether the data is corrupted by stationary or rapidly changing additive noise. In \cite{zt2010}, \emph{a posteriori} SNR weighted energy distance is used as the key speech feature for VAD and has demonstrated state-of-the-art performance in noisy environments, and the rational behind the feature is that the weighting of \emph{a posteriori} SNR makes the weighted distance close to zero in non-speech regions for both clean and noisy speech. The VAD method in \cite{zt2010} first uses the \emph{a posteriori} SNR weighted energy distance for selecting frames in a variable frame rate analysis and then conducts VAD based on the selected frames, and the method assumes the presence of speech within a certain period. While showing state-of-the-art performance, the method in \cite{zt2010} has several drawbacks. First, it makes an implicit assumption of speech presence, which does not always hold, and it is not uncommon that a VAD method assumes the presence of speech in every signal (e.g. a signal file in a database) \cite{VQVAD}, \cite{Sadjadi2013}. Secondly, the process of making VAD decisions through selecting frames first is cumbersome and suboptimal. Finally, estimating \emph{a posteriori} SNR weighted energy distance is prone to noise. In the present work, we therefore propose a VAD method eliminating the assumption of speech presence by using pitch or spectral flatness as an anchor to find potential speech segments, directly using \emph{a posteriori} SNR weighted energy distance without conducting frame selection, and applying speech enhancement.  The proposed method differs from \cite{zt2010} in a number of ways: 1) two-stage denoising is proposed in order to enhance the noise robustness against both rapidly changing and relatively stationary noise, 2) pitch or spectral flatness is applied  to detect high-energy noise segments and as an anchor to find potential speech segments, and 3) a segment based approach is used, in which VAD is conducted in segments, making it easier and more effective in determining a threshold for making decisions since a certain amount of speech and non-speech exists in each segment. The proposed method is called \emph{robust voice activity detection} (rVAD).

Pitch information places an important role in rVAD and it has been used for existing VAD methods as well. For example, \cite{shao2018use} combines pitch continuity with other speech features for detecting voice activity. In \cite{yang2016voice}, long-term pitch divergence is used as the feature for VAD. The big difference here is that rVAD uses pitch as an anchor to locate potential speech regions, rather than as a VAD feature, and in rVAD the actual VAD boundaries are found by using \emph{a posterior} SNR weighted energy distance as the feature. Concerning two-stage denoising, the concept has been applied in the literature although being different. In \cite{zhao2017two}, two-stage denoising is conducted to handle reverberation and additive noise separately in a supervised fashion by training neural networks. A two stage Mel-warped Wiener filter approach is presented in \cite{ETSI2007}. In this work, the two-stage method aims to deal with high-energy noise in the first stage separately. 

Most of the computation in rVAD lies in pitch detection, while the rest part of rVAD is computationally light. Therefore, we present a modified version of rVAD, where the time-consuming pitch detector is replaced by a computationally efficient spectral flatness (SFT) \cite{Peeters2004,Madhu2009,SF_vad2019} detector. 
We call the modified algorithm \emph{rVAD-fast}. We show that rVAD-fast is significantly  ($\approx 10$ times) faster compared with rVAD with moderate degradation on VAD performance, which is beneficial when processing a larger amount of data or used on devices with computational constraints. 
  
We demonstrate the performance of the rVAD method for voice activity detection on two databases consisting of various types of noise: the RATS \cite{RATS} and Aurora-2 \cite{Hirsch2000} databases. We further evaluate its performance for speaker verification on the RATS database as well as on the RedDots database \cite{RedDots} that we corrupt with additive noise. Experiment results show that rVAD is compared favourably with a number of existing methods.  
  
The MATLAB source code of the proposed rVAD method (including rVAD-fast) is made publicly available \footnote{\url{https://github.com/zhenghuatan/rVAD}}, and the Python source code of rVAD-fast is also publicly available \footnote{\url{https://github.com/zhenghuatan/rVADfast}}. It is noted that a slightly different version of the rVAD source code - \emph{rVAD1.0} - has already been made publicly available while no paper has been published to document the rVAD method and a number of studies have used it, covering applications such as voice activity detection in speaker verification \cite{Stefanus2017,Nautsch2016,Dhanush2017}, age and gender identification \cite{shepstone2013audio}, emotion detection and recognition \cite{DBLP:journals/corr/DubeyMM16,Semwal2017, Chorianopoulou2016}, and discovering linguistic structures \cite{lee2014discovering}. A modified version is used for real-time human-robot interaction \cite{zt2017_isorobot}. In addition, we have made the Python source code for training and testing of GMM-UBM and maximum \emph{a posteriori} (MAP) adaptation based speaker verification publicly available \footnote{\url{https://github.com/zhenghuatan/GMM-UBM_MAP_SV}}.

The paper is organized as follows. The two-pass denoising method and the proposed rVAD method are presented in Sections 2 
and \ref{sec:propVAD}, respectively. Section \ref{sec:rVAD-fast} describes rVAD-fast. Experimental results on the Aurora-2 database, the RATS database and the RedDots database are presented in Sections \ref{sec:Aurora-2}, \ref{sec:RATS} and \ref{sec:RedDots}, respectively.
Finally, the paper is concluded in Section \ref{sec:conclusion}. 

\section{Robust VAD in noise}
\label{sec:denoising}

\begin{figure}[H]
\centering\subfigure[\it]{\includegraphics[width=8.2cm,height=4.8cm]{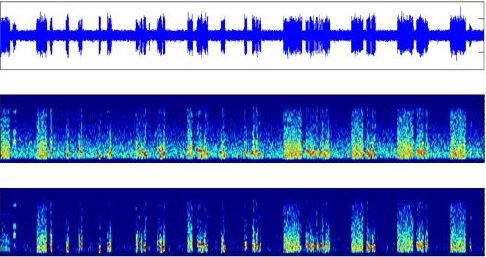}}
\centering\subfigure[\it]{\includegraphics[width=8.2cm,height=4.8cm]{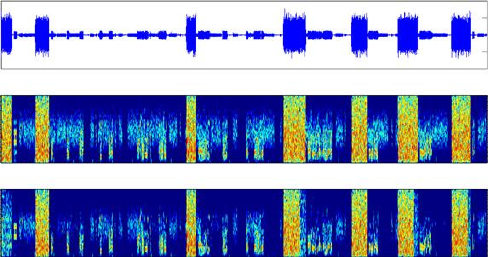}}
\caption{\it Noisy and denoised speech: (a) Noisy speech from Channel A: waveform of noisy speech (the first panel), spectrogram of noisy speech (the second panel) and spectrogram of denoised speech (the third panel); (b) noisy speech from Channel H with the same order of panels as in (a). }
\label{fig:1}
\end{figure}

When robustness is of concern, a major challenge presented by real-world applications is that speech signals often contain both stationary noise as well as burst-like noise. It is very difficult to detect and remove burst-like noise since they have high energy and are rapidly changing and thus hard to estimate. To illustrate this, Fig. \ref{fig:1} shows, from the RATS database \cite{Walker2012}, two examples of noisy speech signals obtained by corrupting clean speech with different types of communication channel noise. The figure depicts waveform and spectrogram of the two noisy speech signals and spectrogram of their corresponding speech signals denoised by the minimum statistics noise estimation (MSNE) \cite{Martin2001a} based spectral subtraction. It is noticed that high-energy noise is largely kept intact after denoising, except for at the beginning of each utterance where the noise estimate is close to the real noise since there is only high-energy noise in the beginning of an utterance. We further did preliminary VAD experiments using a classical statistical method \cite{Sohn1999a} on a few files and the performance is not encouraging (with over 20\% frame error rate). These initial tests show that burst-like noise presents a significant challenge to both denoising and VAD methods.

Due to the very different characteristics of stationary and burst-like noise, the two types of noise require different types of processing. To deal with this problem, we devise a two-pass denoising method, as detailed in the presentation of rVAD in Section \ref{sec:propVAD}. In the first pass, the \emph{a posteriori} SNR weighted energy difference measure \cite{zt2010} is used to detect high-energy segments. If a high-energy segment is detected and it does not contain pitch, the segment is considered non-speech and its samples are set to zero.  

After detecting and denoising high-energy noise segments, in the second pass, a general speech enhancement method is applied to the first-pass-denoised speech signal in order to remove the remaining noise that is relatively more stationary. 
In the classical speech enhancement framework \cite{loizou2013speech}, accurate noise estimation is important and the widely used methods include minimum statistics noise estimation (MSNE) \cite{Martin2001a}, minimum mean-square error (MMSE) \cite{Ephraim1984,Gerkmann2013}, and minima controlled recursive averaging (MCRA) \cite{Cohen2002a}. 
Based on an estimate of additive-noise spectrum, a spectral subtraction (SS) method \cite{Boll1979a} is then used to subtract the estimate from the noisy speech spectrum. 
Another interesting speech enhancement method is the one in the advanced front-end \cite{ETSI2007}, which aims at improving speech recognition in noisy environments. In \cite{Jensen2015a} it is found that the AFE enhancement method outperforms MMSE-based methods for noise-robust speech recognition. 
Recently DNN based speech enhancement methods have also been proposed for improving speech intelligibility \cite{Kolbk2017a}, automatic speech recognition \cite{Wang2017a} and speaker verification \cite{Michelsanti2017, Kolbk2016a}. 


Unsupervised VAD methods often rely on a predefined or adaptive threshold for making VAD decisions, and finding this threshold is a challenging problem in noisy environment. Therefore, it is not uncommon that a VAD method assumes the presence of speech in every signal (e.g. a signal file in a database) that the method is applied upon \cite{VQVAD}, \cite{Sadjadi2013}. This assumption holds for benchmark speech databases, but does not in real-world scenarios where it is possible to have no speech in a long duration. As we know, a speech signal must contain pitch, which motivates us to propose to use pitch (or its replacement speech flatness) as an anchor or indicator for speech presence, namely speech is present in a signal if pitch is detected. This leads to the VAD process in this work: detecting extended pitch segments first and then detecting speech activity within the segments. Extended pitch segment detection plays a key role in rVAD. First, this provides an anchor to locate speech segments. Secondly, it enables to exclude a substantial amount of non-speech, potentially noisy, part from a speech signal. Finally, this results in a segment-based approach in which voice activity detection operates in segments, making it easier and more effective in terms of determining a decision threshold as both speech and non-speech are guaranteed to be present in each segment.

\section{rVAD: an unsupervised segment-based  VAD method}
\label{sec:propVAD}

The block diagram of the proposed rVAD method is shown in Fig. \ref{fig:Fig2}. It consists of the following steps: the first pass denoising (high-energy segment detection, pitch based noise-segment classification, and setting high-energy noise segments to zero), the second pass denoising, extended pitch segment detection, and the actual VAD. These steps are detailed in this section.

The noise-corrupted speech signal is modelled using the additive noise signal model as
\begin{equation}
x(n) = s(n) + v(n)
\end{equation}
                                                                                                                             
\noindent where $x(n)$ and $s(n)$ represent the noisy and clean speech sample at time $n$, respectively, and $v(n)$ the sample of additive noise at time $n$. The signal $x(n)$ is first filtered by a high-pass filter to remove the DC component and low frequency noise. A first-order high-pass filter with a cutoff frequency of 60 Hz is applied to remove low-frequency noise. For simplicity, this is not included in the equations of this paper. 

To conduct short-time speech analysis, the signal is partitioned into frames of 25 ms in length with a frame shift of 10 ms, without using pre-emphasis or a hamming window unless stated otherwise.

\subsection{The first pass denoising }
In the first pass, high-energy segments are detected by using an \emph{a posteriori} SNR weighted energy difference measure \cite{zt2010} as follows:
\begin{itemize}
\item[{\bf (a)}] Calculate the \emph{a posteriori} SNR weighted energy difference of two consecutive frames as
\begin{equation}
d(m) = \sqrt[]{|e(m)-e(m-1)|\ast \max(SNR_{post}(m),0)} \label{eq:dt}
\end{equation} 
where $m$ is the frame index, $e(m)$ the energy of the $m^{th}$ frame of noisy speech $x(n)$, and $SNR_{post}(m)$ is \emph{a posteriori} SNR that is calculated as the logarithmic ratio of $e(m)$ to the estimated energy of the $m^{th}$ frame of noise signal $v(n)$:
\begin{equation}
SNR_{post}(m)=10\ast log_{10}\frac{e(m)}{\tilde{e}_v(m)} \label{eq:postSNR}
\end{equation}

In Eq.(\ref{eq:dt}), the square root is taken to reduce the dynamic range, which differs from \cite{zt2010} where the square root is not applied. Energy is calculated as the sum of the squares of all samples in a frame. The noise energy $\tilde{e}_v(m)$ is estimated as follows. First, the noisy speech signal $x(n)$ is partitioned into super-segments of 200 frames each (about 2s): $x(p)=s(p)+v(p)$, $p=1,\ldots, P$, where $P$ is the number of super-segments in an utterance. For each super-segment $x(p)$, the noise energy $e_v(p)$ is calculated as the energy of the frame ranked at $10\%$ of lowest energy within the super-segment. Thereafter, the noise energy $\tilde{e}_v(p)$ is calculated as the smoothed version of $e_v(p)$ with a forgetting factor of 0.9 as follows:
\begin{equation}
\tilde{e}_v(p)= 0.9 \ast \tilde{e}_v(p-1) + 0.1 \ast e_v(p).
\end{equation}
Noise energy of the $m^{th}$ frame, $\tilde{e}_v(m)$, takes the energy value $\tilde{e}_v(p)$ of the $p^{th}$ super-segment which the   $m^{th}$ frame belongs to.

\begin{figure}[H]
\centering\includegraphics[]{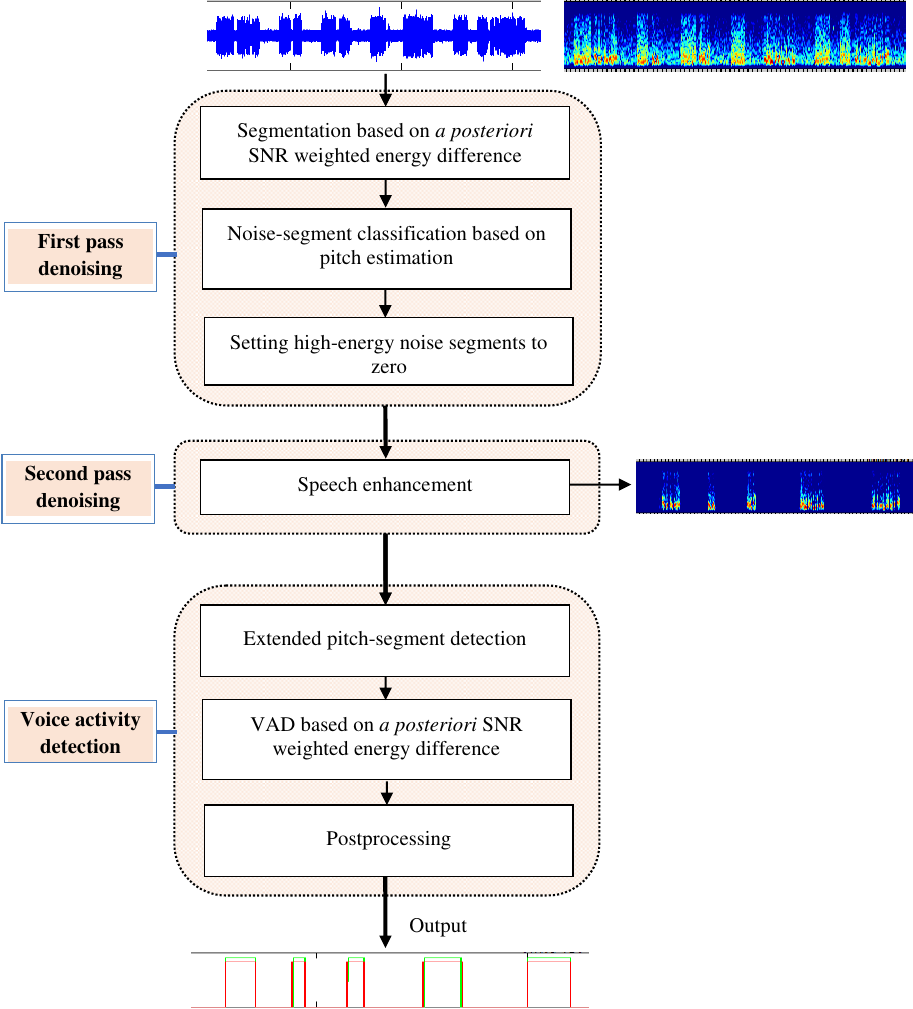}
\caption{\it Block diagram of the rVAD method.}
\label{fig:Fig2}
\end{figure}

\item[{\bf (b)}] Central-smooth the \emph{a posteriori} SNR weighted energy difference,
\begin{equation}
\bar{d}(m) = \frac{1}{2N+1}\sum_{i=-N}^{N} d(m+i) \label{eq:smDt}
\end{equation}
where $N=18$.

\item[{\bf (c)}] Classify a frame as a high-energy frame if $\bar{d}(m)$ is greater than a threshold $\theta_{he}(m)$. For each super-segment $p$ (containing 200 frames), $\theta_{he}(p)$ is computed as follows: 
\begin{equation}
\theta_{he}(p)= \alpha \ast max \big \{ e((p-1)*200+1), \ldots, e(m) \ldots, e(p*200)\big \}
\end{equation}
where $\alpha=0.25$. 
$\theta_{he}(m)$ takes the threshold value $\theta_{he}(p)$ of the $p^{th}$ super-segment  which the $m^{th}$ frame  belongs to. 
Alternatively, the threshold can be calculated recursively using a forgetting factor.

\item[{\bf (d)}] Consecutive high-energy frames are grouped together to form high-energy segments. 
\item[{\bf (e)}] Within a high-energy segment, if no more than two pitch frames are found, the segment is classified as noise, and the samples of the segment are set to zero. 
\end{itemize}

Motivation of this pass is two-fold: first, to avoid overestimating noise due to the burst-like noise when applying a noise estimator  in the second pass denoising and secondly, to detect and denoise high-energy non-speech parts, which are otherwise difficult for conventional denoising and VAD methods to deal with.

\subsection{The second pass denoising}
Any speech enhancement method is applicable for the second pass denoising. Three spectral subtraction based speech enhancement methods are considered in this work, and they rely on different noise estimation approaches: MMSE \cite{Gerkmann2013}, MSNE \cite{Martin2001a}, and a modified version of MSNE (MSNE-mod).

The unbiased noise power estimation in the conventional MSNE can be expressed as,
\begin{equation}
\hat{\lambda}_v(m,k) = B_{min}\ast min\big \{P(m,k), P(m-1,k), \ldots,P(m-l,k) \big \}
\end{equation}
where $m$ is the frame index, $k$ the frequency bin index, $B_{min}(m,k)$ the bias compensation factor, $P(m,k)$   the recursively smoothed periodogram, and $l$ the length of the finite window for searching the minimum. 

In the proposed MSNE-mod, noise estimate   $\hat{\lambda}_v(m,k)$ is not updated during the detected high-energy noise segments (which are set to zero).  Besides, if more than half of the energy is located within the first 7 frequency bins ($< 217\; Hz$), the values of the $7$ frequency bins 
are set to zero to further remove low-frequency noise in addition to the use of the first-order high-pass filter mentioned earlier.

\subsection{Extended pitch segment detection}
The VAD algorithm is based on the denoised speech and the pitch information generated from the previous steps. The fundamental assumption is that all speech segments should contain a number of speech frames with pitch. In this algorithm, pitch frames are first grouped into pitch segments, which are then extended from both ends by $60$ frames ($600$ms), based on speech statistics, in order to include voiced sounds, unvoiced sounds and potentially non-speech parts. This strategy is taken for the following reasons: 1) pitch information is already extracted in the previous steps, 2) pitch is used as an anchor to trigger the VAD process and 3) many frames can be potentially discarded if an utterance contains a large portion of non-speech segments, which are also non-pitch segments. \\

\subsection{Voice activity detection}
The \emph{a posteriori} SNR weighted energy difference is applied now to each extended pitch segment  to make VAD decision as follows:
\begin{itemize}
\item[{\bf(a)}] Calculate \emph{a posteriori} SNR weighted energy difference $d^{'}(m)$ according to Eq. (\ref{eq:dt}). To calculate $SNR_{post}(m)$ using Eq. (\ref{eq:postSNR}), the noise energy $\tilde{e}^{'}_v(m)$ is estimated as the energy ranked at $10\%$ of lowest frame energy within the extended pitch segment.

\item[{\bf(b)}]	Central-smooth the \emph{a posteriori} SNR weighted energy difference with $N=18$ as in Eq.(\ref{eq:smDt}), resulting in $\bar{d}\:^{'}(m)$. 

\item[{\bf(c)}]Classify a frame as speech if $\bar{d}\:^{'}(m)$  is greater than the threshold ($\theta_{vad}$)  
\begin{equation}
\theta_{vad} = \beta \ast \frac{1}{L}\sum_{j=1}^{L} \bar{d}\:^{'}(j) \label{eq:vadthr}
\end{equation}
where $L$ is the total number of frames with pitch in the extended pitch segment and the default value for $\beta$ is set to 0.4. 

\item[ {\bf(d)}] Apply post-processing. The assumptions here are speech frames should not be too far away from its closest pitch frame, and within a speech segment, there should be a certain number of speech frames without pitch. While it is possible to gain improvement by analysing noisy speech as well, we analyse only a few clean speech files that are not included in any experiments in this paper and then derive and apply the following rules. First, frames that are $33$ frames away from the pitch segment to the left and  $47$ frames away to the right are classified as non-speech, regardless the VAD results above, which covers 95\% of the cases based on the a few speech files. On the other hand, frames that are within  $5$ frames to the left and  $12$ frames to the right of the pitch segments are classified as speech, again regardless the VAD results, which leaves out 5\% of the cases based on the a few speech files. The concept is sort of similar to that of hangover schemes used in some VAD methods. Furthermore, segments with energy below 0.05 times the overall energy is removed. 

\end{itemize}

\section{rVAD-fast based on spectral flatness}
\label{sec:rVAD-fast}
As the computational complexity of pitch detection is relatively high, we investigate alternative measures to pitch, for example, spectral flatness (SFT) \cite{Peeters2004,Madhu2009,Johnston1988}. Our primitive study shows that SFT is a good indicator to tell whether or not there is pitch in a speech frame. Replacing the pitch detector by a simple SFT based voiced/unvoiced speech detector, leads to a more computationally efficient algorithm called rVAD-fast.
To extract the SFT feature, a hamming window is first applied to a speech frame before taking  short-time Fourier transform (STFT). After STFT, the signal is represented in the spectral domain as
\begin{equation}
X(m,k) = S(m,k) + V(m,k).
\end{equation}
\noindent 
Thereafter, SFT is calculated as 
\begin{equation}
SFT (m) = \frac{exp(\frac{1}{K} \sum_{k=0}^{K -1} ln\; |X(m,k)|)} {\frac{1}{K} \sum_{k=0}^{K -1} |X(m,k)|}
\end{equation}

\noindent where $|X(m,k)|$  denotes the magnitude spectrum of $k^{th}$ frequency bin for the $m^{th}$ frame, and $K$ is the total number of frequency bins. As SFT is used as a replacement for pitch in this work, SFT values are compared against a predefined threshold  $\theta_{sft}$ to decide whether their corresponding frame is voiced or unvoiced. 

\begin{figure}[H] \label{fig:ft}
\centering{\subfigure[\it]{\includegraphics[width=12.2cm,height=5.5cm]{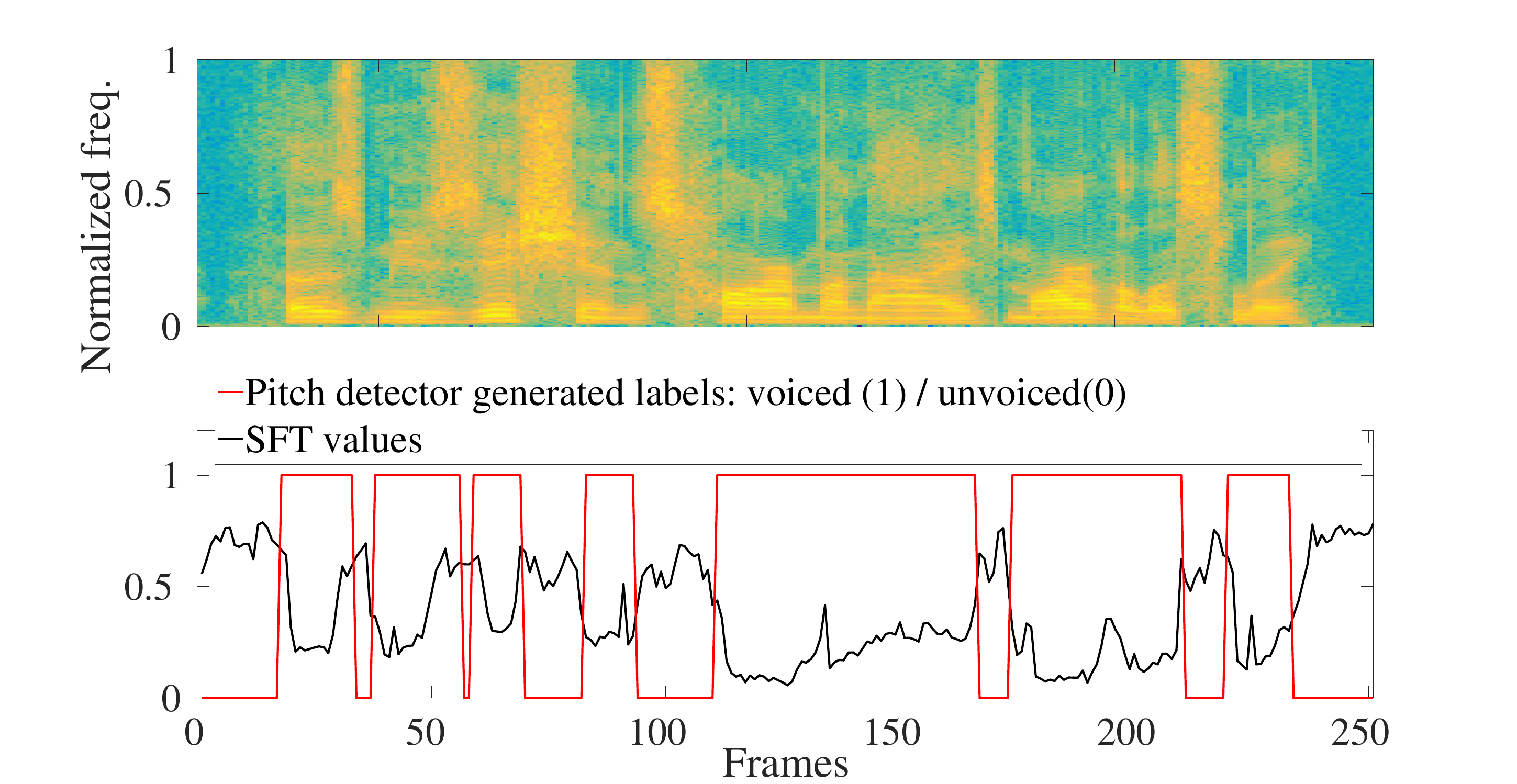}\label{fig:ft_timit}}} \\
\centering{\subfigure[\it]{\includegraphics[width=12.2cm,height=5.5cm]{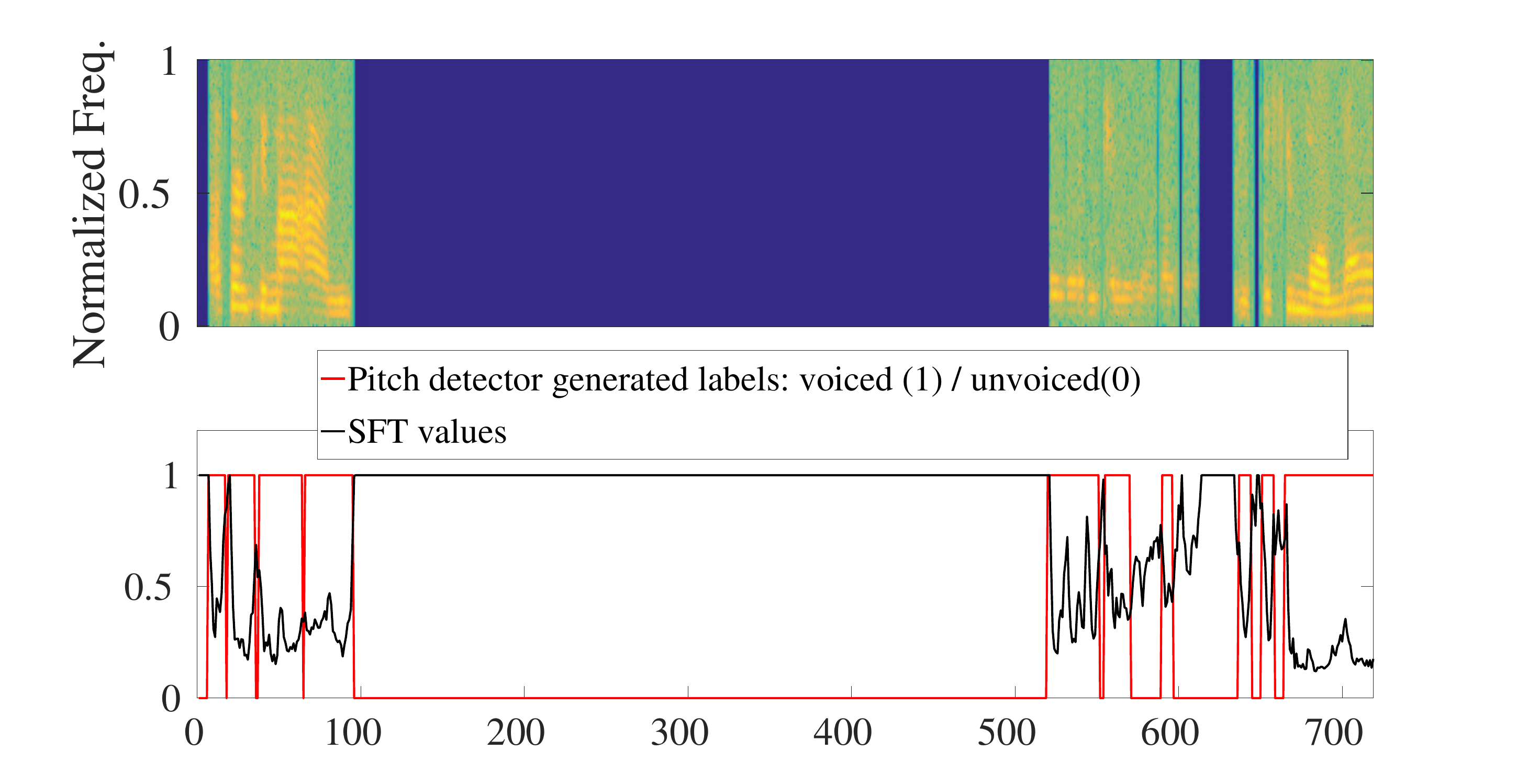}
\label{fig:ft_nist16}}}
\caption{\it  Spectrogram, pitch labels (1 for pitch and  0 for no pitch) and SFT values of a speech signal from (a) TIMIT (b) the NIST 2016 SRE evaluation set.}
\end{figure}

Figures \ref{fig:ft_timit} and \ref{fig:ft_nist16} illustrate spectrogram, pitch labels (1 for pitch and  0 for no pitch) and SFT values of a speech signal from TIMIT (clean) and those of a signal from the NIST 2016 SRE evaluation (noisy), respectively. The figures show that if we choose a $\theta_{sft}$ value of $0.5$ (i.e. if SFT $\leq 0.5$, a frame is said to contain pitch), the labels  generated by  SFT are close to those generated by the pitch detector.

We extensively studied the effect of different threshold values $\theta_{sft}$ on the performance of SFT as a replacement of the pitch detector, which will not be detailed in this paper. Briefly, we compared the output of the SFT detector of different values of $\theta_{sft}$ with the output of the pitch detector on large number of utterances from various databases including NIST 2016 SRE (evaluation set) \cite{NIST2016SRE}, TIMIT \cite{Timit}, RSR2015 \cite{RSR2015}, RedDots \cite{RedDots}, ASVspoof2015 \cite{wu2015} and  noisy versions (car, street, market, and white) of ASVspoof2015. It is observed that a threshold value of $0.5$ gives the best match between SFT and pitch, and this value is used for experiments in this paper.

\section{Experiments on the Aurora-2 database} 
\label{sec:Aurora-2}
To evaluate the performance of the proposed rVAD method, experiments are conducted on a number of databases and for different tasks. In this section, we compare rVAD with ten existing VAD methods (both supervised and unsupervised) in terms of VAD performance on the test sets of the Aurora-2 database. Aurora-2 \cite{Hirsch2000} has three test sets A, B and C, all of which contain both clean and noisy speech signals. The noisy signals in Set A are generated from clean data by using a filter with a G.712 characteristic and mixing four noise types including subway, babble, car and exhibition with SNR values ranging across 20 dB, 15 dB, 10 dB, 5 dB, 0 dB, and -5 dB. Set B is created similarly with the only difference being that the types of noise are restaurant, street, airport and train station. In Set C, clean speech is corrupted by subway and street noise, in addition to a Motorola integrated radio systems (MIRS) characteristic filter being applied instead of that of G.712.

The reference VAD labels for the Aurora-2 database are generated with the HTK recognizer \cite{htkbook} which is trained using the training set of Aurora-2. Whole word models are created for all digits. Each of the whole word digit models has $16$ HMM states with three Gaussian mixtures per state. The silence model has three HMM states with six Gaussian mixtures per state. A one state short pause model is tied to the second state of the silence model. The speech feature consists of $12$ MFCC coefficients, logarithmic energy as well as their corresponding $\Delta$ and $\Delta\Delta$ components. In \cite{kraljevski2015comparison}, it is confirmed that forced-alignment speech recognition is able to provide accurate and consistent VAD labels, matching closely transcriptions made by an expert labeler and being better than most of those made by non-expert labelers. The generated  reference VAD labels are made publicly available 
\footnote{\url{https://github.com/zhenghuatan/rVAD}}.

Several metrics are used to characterize VAD performance. The frame error rate (FER) is defined as

\begin{equation}
FER =100\ast \frac{\#  \; \big \{false \;rejection\;frames  \; + \;false \;alarm\; frames \big \}  }{\#\; total \;frames}
\end{equation}

False alarm rate $P_{fa}$ is the percentage of non-speech frames being misclassified as speech and miss rate $P_{miss}$ is the percentage of speech frames being misclassified as non-speech. Detection cost function (DCF) is defined as 
\begin{equation}
DCF = (1-\gamma)\ast P_{miss} + \gamma\ast P_{fa}
\end{equation}
where the weight $\gamma$ is equal to $0.25$, which penalizes missed speech frames more heavily.  


Through out this paper, the default configuration for rVAD is the one that includes two-pass denoising (with the second being MSNE) and the post-processing, which is also the default configuration in the released rVAD source code, and this configuration, shown in italic font in tables, is used for all experiments in this paper unless stated otherwise. We do not change the settings and parameters while testing rVAD on different datasets so as to evaluate its generalization ability, unless a test is specifically for assessing the effects of changing settings, e.g. the threshold $\beta$ in Eq. (\ref{eq:vadthr}).   

For experiments conducted in this paper, pitch extraction is realized using the pitch estimator in \cite{Gonzalez2011}, unless stated otherwise. It has been experimentally shown that using Praat pitch extraction \cite{Boersma2009} gives almost the same VAD results.  

\subsection{Comparison with referenced methods and evaluation of different configurations}

Table \ref{table:table1} presents the VAD results averaged over the three test sets of Aurora-2, which accounts for $70070$ speech files, for various methods. VQVAD \cite{VQVAD} first applies a speech enhancement method and an energy VAD to a testing utterance in order to automatically label a small subset of MFCCs as speech or non-speech; afterwards, these MFCCs are used to train speech and non-speech codebooks using k-means and all the frames in the utterance are labeled using nearest-neighbor classification.
Sohn \emph{et al.} VAD \cite{Sohn1999a} is  an unsupervised VAD method based on a statistical likelihood ratio test, which uses Gaussian models to represent the distributions of speech and non-speech energies in individual frequency bands. The method also uses a hangover scheme. The VoiceBox toolkit \footnote{\url{http://www.ee.ic.ac.uk/hp/staff/dmb/voicebox/voicebox.html}} implemented version of Sohn \emph{et al.} VAD  is used in this study. 
\emph{Kaldi} VAD is the the Kaldi toolkit's \cite{Povey_ASRU2011} energy based VAD for which we use the default parameters (-vad-energy-threshold=5.5, --vad-energy-mean-scale=0.5) as included in the SRE16 script. Note that Kaldi is a widely used open source software for speaker and speech recognition. 

Results for the G.729, G.723.1 and MFB VAD methods are cited from \cite{Vlaj2005a}, results for the LTSV and GMM-NLSM methods are from \cite{Vlaj2012}, and results for the DSR-AFE and variable frame rate (VFR) methods are from \cite{zt2010}.  The comparison in this table is conducted in terms of frame error rate (FER) since results of LTSV and GMM-NSLM are only available in terms of FER. Note that the identical experimental settings and labels are used across \cite{zt2010, Vlaj2005a, Vlaj2012} and the present work, so the comparison is valid. 

The results in Table \ref{table:table1} clearly show that rVAD with the default configuration gives significantly lower average FER than those of the compared VAD methods. It outperforms all referenced VAD methods, with big margins, under all SNR levels including the Clean condition. The closest one is the VFR VAD \cite{zt2010}, which is our previous work that also uses \emph{a posteriori} SNR weighted energy distance as the feature for VAD decision. The GMM-NLSM \cite{Vlaj2012} VAD provides good performance as well, but still with a 3\% (absolute) higher FER as compared with rVAD, and furthermore it should be noted that GMM-NLSM is a \emph{supervised} VAD where the GMMs are trained using multicondition training data of the Aurora 2 database. The next one in line is the VAD method in the DSR AFE frontend \cite{ETSI2007}, which is an unsupervised VAD and gives a more than 5\% (absolute) higher FER than that of rVAD. DSR AFE performs well under low SNRs, for example at 0dB and -5dB, with FERs close to those of rVAD; from 5dB and above, however, its performance is far below that of rVAD. The remaining compared methods have even much higher FERs. For example, VQVAD\cite{VQVAD} gives a 36\% FER, as compared with a 11\% FER achieved by rVAD.

\begin{table}[H]
\caption{\it Comparison of rVAD (with the default configuration) with other methods on the test datasets of Aurora-2.} 
\footnotesize{
\begin{center}
\begin{tabular}{|l|lllllll|l|}\cline{1-9}
 Methods          &   \multicolumn{7}{c|}{FER (\%)} &  Avg.\\ 
                  &  Clean   &20 dB & 15 dB & 10 dB & 5dB & 0 dB & -5 dB & FER  \\ \hline
 VQVAD\cite{VQVAD}            &17.26   & 33.64  & 36.34  & 38.77  & 41.05 & 43.03  &46.07   & 36.59 \\
G.729 \cite{ITU1996a}                    & 12.84  & 24.53  & 26.13  & 27.38  & 29.13& 32.23 & 35.21  & 26.78 \\
Sohn et al. \cite{Sohn1999a}                  & 17.37	&20.16	& 21.97	  &24.48& 27.96	& 33.12	& 39.76 &  26.40\\
Kaldi Energy VAD \cite{Povey_ASRU2011}            &9.88	 &26.54	  & 26.61  & 26.62	& 26.62 &26.62   &26.62   & 24.22 \\
G.723.1 \cite{ITU1996b}                      & 19.45  &21.31   & 23.29  & 24.44  & 26.30& 26.56 & 28.58  & 23.99 \\
MFB \cite{Vlaj2005a}                        & 6.92   & 15.39  & 17.70  & 20.12  & 22.75& 26.16 & 31.09  & 20.02 \\
LTSV \cite{Ghosh2011}                          & 9.50   & 15.90  & 16.80  & 18.20  & 21.00& 26.10 & 28.80  & 19.50 \\
DSR AFE \cite{ETSI2007}                   & 18.41  & 15.16  & 14.96  & 14.59  & 14.54& 15.62 &  22.08 & 16.48 \\
GMM-NLSM (supervised) \cite{Vlaj2012}                   & 10.95  & 11.20  & 11.43  &11.73   & 13.35 &17.44 & 23.52  & 14.23 \\ 
VFR \cite{zt2010}                            & 8.10   & 8.30   & 9.00   & 10.60  & 13.50 & 19.50 & 28.20 & 13.90 \\ 
\emph{rVAD (default)}       &{\bf6.90}  &{\bf7.30} &{\bf7.64}  &{\bf8.43} & {\bf11.09} &{\bf16.01} &{\bf21.48} &{\bf11.26}\\

\hline
\end{tabular} 
\end{center}
}
\label{table:table1}
\end{table}	

Table \ref{table:table1b} compares the results of various configurations for rVAD. It is shown that when evaluating on the Aurora-2 database, the first-pass denoising of rVAD does not make difference, for which the reason is that burst-like high-energy noise prominently present in the RATS database has much less presence in Aurora-2 while the aim of the first pass denoising is to remove burst-like noise. 
The second pass denoising with MSNE or MMSE is able to boost the performance of rVAD and there is almost no performance difference between the two enhancement methods. MSNE-mod, however, does not perform as well as MSNE, for which the reason is that MSNE-mod is tailored to the special noise characteristics of the RATS database. 
The postprocessing step is also shown to be important for improving the performance of rVAD. 

\begin{table}[H]
\caption{\it Comparison of rVAD (with different configurations) with other methods on the test datasets of Aurora-2.} 
\footnotesize{
\begin{center}
\begin{tabular}{|l|ll|lllllll|l|}\cline{1-11}
 Methods          &  \multicolumn{2}{c|}{Denoising}   & \multicolumn{7}{c|}{FER (\%)} &  Avg.\\ 
                  &  1st-pass  & 2nd-pass                        &Clean   &20 dB & 15 dB & 10 dB & 5dB & 0 dB & -5 dB & FER  \\ \hline
\emph{rVAD (default)}   & $\checkmark$ & MSNE                           &6.90  &7.30 &7.64      &{\bf8.43} &11.09 &16.01 &21.48 &11.26\\
(two-pass denoising)     &$\checkmark$ & MMSE                           &6.96  &7.35 &7.75	    &8.58      &11.02 &15.93 &21.55 &11.30\\
                      &$\checkmark$ &MSNE-mod                           &6.86  &7.07 &{\bf7.44}	&8.68      &13.13 &19.73 &25.56 &12.63\\ 

                                            &         &                             &       &        &        &         &        &       &  & \\
rVAD (w/o denoising)   &$\times$  & $\times$                    & 6.89 &7.37 &7.69  &8.81 &13.07 &19.05 &23.23 &12.30  \\
rVAD (w/o 1st-pass  &$\times$     & MSNE                        & 6.89 &7.30 & 7.64 &8.45 &11.01 &15.88 &21.44 &{\bf11.23}\\
\hspace{1cm} denosing)   & $\times$  & MMSE                     &6.95  &7.35 &7.76  &8.60 &11.00 & 15.85 &21.51 & 11.28 \\
      & $\times$ & MSNE-mod                                     &{\bf6.85}  &{\bf7.06} &{\bf7.44}	&8.66 &13.07  &19.59 &25.52   &12.59\\ 
      &          &                              &       &        &        &         &        &       &  & \\
rVAD (w/o postproc)  & $\checkmark$ & MSNE       &8.82  &9.53 &9.78  &10.05&11.32 &15.47& 21.59 & 12.37\\

\hline
\end{tabular} 
\end{center}
}
\label{table:table1b}
\end{table}		

Overall, the experimental results demonstrate the effectiveness of the rVAD method on the Aurora-2 database, which is a very different database from the RATS database that rVAD was originally devised for. This confirms the good generalization ability of rVAD. 

\subsection{Sensitivity of rVAD with changing threshold $\beta$ on VAD performance}
In rVAD, $\beta$ in Eq. (\ref{eq:vadthr}) is an important thresholding parameter that controls the aggressive level of rVAD; the larger the value of $\beta$, the higher the aggressive level. 
Table \ref{table:aurora_res_sensitivity} presents VAD results, in terms of false rejections rate $P_{miss}$, false alarms rate $P_{fa}$ and FER, of rVAD with various $\beta$ values. Apart from varying $\beta$ values, the default configuration of rVAD is used. For small $\beta$ values, $P_{miss}$ is small while $P_{fa}$ is large and vice versa. As of FER, a threshold value of 0.4 gives the best performance while a change of $\pm$ 0.1 in $\beta$ only marginally change the performance. The results confirm the stability of rVAD with respect to changing this threshold.

In Table \ref{table:aurora_res_sensitivity}, we additionally compare rVAD with several other methods now also in terms of $P_{miss}$ and $P_{fa}$ while in Table \ref{table:table1} only FER performance is compared. This allows us to compare rVAD with other methods in terms of $P_{fa}$, while fixing $P_{miss}$ or the other way around. Table \ref{table:aurora_res_sensitivity} clearly shows that rVAD outperforms the referenced methods with big margins.   

\begin{table}[H]
\caption{\it Performance of rVAD with various threshold $\beta$ values in comparison with several referenced methods on the Aurora-2 database.}
\begin{center}
\begin{tabular}{|lcccc|}\cline{1-5}
 Method                & Threshold &  $P_{miss}(\%)$ & $P_{fa}(\%)$ &Avg. FER \\              
                       & ($\beta$) &             &    &     (\%)    \\  \hline
                      & 0.1       &  1.21     & 52.90  & 14.97  \\
rVAD                  & 0.2       &  2.16     & 42.43   &12.87 \\

  & 0.3       &  3.21     & 35.11   &11.70 \\
        &   0.4 (default)& 4.86   & 28.91   & 11.26 \\
        &             0.5         & 7.54    & 23.59   &11.81 \\
        &             0.6         & 11.41    & 19.03   &13.44 \\
        &             0.7         & 16.48    & 15.25   &16.15 \\ 
                      &            &             &         & \\
VQVAD \cite{VQVAD}             &   -   &   7.02            &     47.32     & 36.59  \\
Sohn et al. \cite{Sohn1999a}         &      -      &  17.31    & 51.48        & 26.40 \\
Kaldi Energy VAD \cite{Povey_ASRU2011}      &   -  &  1.53     & 86.75    &    24.22 \\  
DSR AFE \cite{ETSI2007}  &      -      &  3.32          & 56.61         & 16.48\\ \hline
\end{tabular}
\end{center}
\label{table:aurora_res_sensitivity}
\end{table}	                    

\subsection{Evaluation of rVAD-fast}

Table \ref{table:tableSFT} compares the performance of the rVAD-fast method with rVAD on the test datasets of Aurora-2  in terms of FER and processing time. It can be seen  that rVAD-fast is approximately an order of magnitude faster than rVAD at the cost of moderate performance degradation. This presents an advantage when processing a large number of speech files or running on low-resource devices. For measuring the processing time, the algorithms were run on a desktop computer with Intel(R) Core(TM) i7-4790 CPU @ 3.60GHz and 16 GB RAM, and we measured the CPU time. 

\begin{table}[H]
\caption{\it Comparison of rVAD-fast with rVAD for voice activity detection on the test datasets of Aurora-2  in terms of FER and average CPU processing  time}
\begin{center}
\footnotesize{
\begin{tabular}{|l|lllllll|l|l|c|}\cline{1-11}
Methods                              &  \multicolumn{8}{c|}{FER (\%)}                         & Avg. time  & Times as fast \\
                         & Clean   & 20 dB  & 15 dB  & 10 dB  & 5 dB   & 0 dB  & -5 dB & Avg. & sec/file & rVAD-fast \\ \hline
 \emph{rVAD (MSNE, default)}             & 6.90	   & 7.30   &7.64    &	8.43  & 11.09  & 16.01 &21.48  & 11.26 & 0.1253  &  \\
 rVAD-fast(MSNE)         & 7.25	   & 8.75   &9.15    &10.07   &12.37   &17.35  &25.18  &12.87  &  0.0085 & $\approx14\times$ \\ 
                         &         &        &        &        &        &       &       &       &         &    \\
 rVAD(MMSE)              & 6.96    &7.35    &7.75	&8.58	  &11.02   &15.93 &21.55   &11.30  & 0.1181  & \\
 rVAD-fast(MMSE)         & 7.26	   & 8.78   &9.22   &10.20    &12.61   &17.71 &25.95   &13.10  & 0.0068  &  $\approx17\times$\\ 
                         &         &        &       &         &        &      &        &       &         &   \\ 
 rVAD(MSNE-mod)         & 6.86    &7.07	& 7.44	&8.68	  &13.13  &19.73  &25.56   &12.63  &   0.1206 &  \\ 
rVAD-fast(MSNE-mod)     & 7.25	   &8.71    &9.30   &11.29    &15.31  &21.01  &30.40   &14.75  &   0.0135 & $\approx 9\times$\\ \hline 
\end{tabular}
}
\end{center}
\label{table:tableSFT}
\end{table}

\section{Experiments on the RATS database}
\label{sec:RATS}
In this section, rVAD is evaluated against nine different VAD methods (both supervised and unsupervised) on the RATS database \cite{RATS,LDC_rats} and it consists of audio recordings selected from several existing and new data sources. All recordings are re-transmitted through $8$ different noisy communication channels, labelled by the letters A through H. The database is for evaluating methods such as voice activity detection, speaker identification, language identification and keyword spotting. 

\subsection{Illustrative VAD and denoising results of rVAD}
\label{sec:illusresults}
\begin{figure}[H] 
\centering{\subfigure[\it]{\includegraphics[width=8.2cm,height=4.0cm]{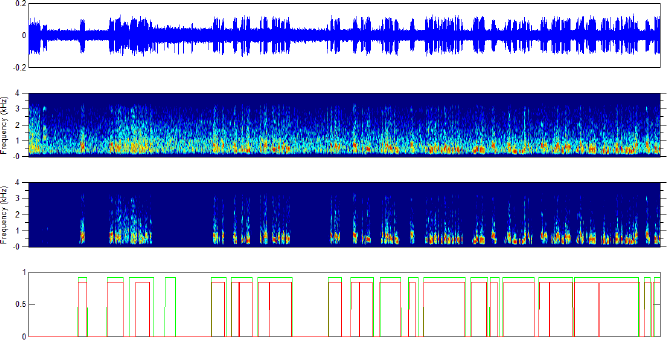}\label{fig:Fig4c}}} \\
\centering{\subfigure[\it]{\includegraphics[width=8.2cm,height=4.0cm]{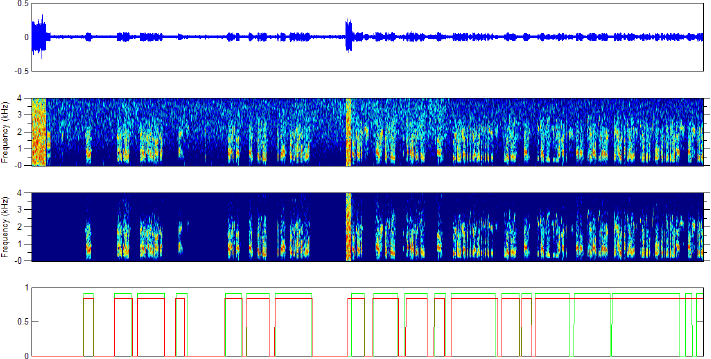}
\label{fig:Fig4d}}}
\caption{\it  Illustrative results of the proposed rVAD method on speech signals from the RATS database: (a) noisy speech from Channel A with four panels presenting waveform of the original signal, its spectrogram, spectrogram of the denoised signal and VAD results where green and red colours represent true labels and algorithm outputs, respectively; b) noisy speech from Channel H with the same order of panels as in (a).}
\label{fig:Fig3}
\end{figure}

Figure \ref{fig:Fig3} illustrates results of rVAD for two speech recordings taken from the RATS database. It shows that rVAD is able to remove both stationary and burst-like noise and performs well in terms of VAD.

\subsection{VAD results}
\label{sec:VAD_results}

Data used for this evaluation contains $8$x$25=200$ speech files randomly selected from the RATS database and it amounts to approximately $44$ hours of speech in total, which is a sufficient quantity for evaluating VAD methods. 
Table \ref{table:rats_res_200} compares rVAD of default configuration with other VAD methods. 

Table \ref{table:rats_res_200} shows that rVAD gives substantially lower FER compared with Sohn \emph{et al.},  Kaldi \cite{Povey_ASRU2011}, VQVAD \cite{VQVAD} and SSGMM \cite{sholokhov2018semi} VAD methods. SSGMM \cite{sholokhov2018semi} is neighter included in the experiments for the Aurora-2 database nor for the RedDots database, as SSGMM does not work for short utterances (due to the data sparsity issue caused by short utterances when training GMMs), as also stated in \cite{sholokhov2018semi}, and it produces empty label files for Aurora-2 or RedDots utterances. 
 Kaldi VAD provides the lowest value of $P_{miss}$, but with significantly higher $P_{fa}$ and FER values. $P_{fa}=86\%$ indicates that Kaldi VAD classifies most frames as speech. Apparently, simple energy-threshold based VAD methods do not work well under highly noisy environments. VQVAD, a widely used and well-performing VAD method in the speaker verification domain, provides as high as 36\% FER. It is interesting to note that VQVAD and Sohn \emph{et al.} achieve 36\% and 26\% FER, respectively, on both the Aurora-2 database and the RATS database, as shown in Tables \ref{table:table1} and \ref{table:rats_res_200}.

\begin{table}[H]
\caption{\it Comparison of VAD performance of the proposed rVAD of default configuration with other methods on the RATS database.}
\begin{center}
\begin{tabular}{|l|lll|}\cline{1-4}
 Methods            &       $P_{miss}(\%)$ & $P_{fa(\%)}$ & FER (\%) \\ \hline
 Kaldi Energy VAD \cite{Povey_ASRU2011}     & 3.10  & 85.98    &51.62    \\
 DSR AFE \cite{ETSI2007}                    & 4.17  & 76.24    & 46.36 \\
 VQVAD\cite{VQVAD}             &  41.65  & 32.85 & 36.49  \\
 Sohn et al. \cite{Sohn1999a}             & 30.36        & 24.51    & 26.93   \\
 SSGMM\cite{sholokhov2018semi}        &  30.36       & 13.24     & 20.33 \\
 \emph{rVAD (default)}          & 17.00        &5.52	& {\bf 10.27} \\ \hline                        
\end{tabular}
\end{center}
\label{table:rats_res_200}
\end{table}

Concerning the different configurations of rVAD, Table \ref{table:rats-res-denoising} shows the first-pass denoising improves the performance slightly. The modified MSNE, which is tailored towards the RATS database, boosts the VAD performance modestly while MSNE and MMSE does not. The best performance is given by first-pass denoising and modified MSNE based second-pass denoising. Overall, the improvement brought by denoising is modest, for which the reason could be that the voice activity detection is conducted on the basis of the extended pitch segments in which the non-speech segments that do not contain pitch have been excluded during the process of detecting pitch segments. 

\begin{table}[H]
\caption{\it Comparison of VAD performance of the proposed rVAD of different configurations on the RATS database.}
\begin{center}
\begin{tabular}{|l|ll|lll|}\cline{1-6}
 Methods      & \multicolumn{2}{c|}{Denoising}   & $P_{miss}(\%)$ & $P_{fa}(\%)$ & FER (\%) \\ \cline{2-3}
                            & 1st-pass       & 2nd-pass                         &            &          &     \\  \hline
              
 \emph{rVAD (default)}       & $\checkmark$  & MSNE                             & 17.00        &5.52	& 10.27 \\
 (with two-pass denoising)   &$\checkmark$   & MMSE                             & 16.67	       &5.84	&10.32\\
                             &$\checkmark$   & MSNE-mod                         & 17.63        & 4.82	& {\bf 10.12} \\ 
 &             &              &               &             &        \\
 rVAD (w/o denoising)        &$\times$       & $\times$                         & 16.30        &6.59	&10.61 \\
 rVAD (w/o 1st-pass          &$\times$       & MSNE                             & 17.88	       &5.63	&10.71 \\
 \hspace{1cm} denoising)     &$\times$       & MMSE                             & 18.21    	   &5.56	&10.80 \\ 
                             &$\times$       & MSNE-mod                         & 17.52	       &5.16	& 10.28 \\
                        &             &                        &               &              &  \\
rVAD (w/o postproc)          &$\checkmark$   & MSNE                            & 18.07         & 8.29   & 12.34  \\ \hline                        
\end{tabular}
\end{center}
\label{table:rats-res-denoising}
\end{table}

We study the sensitivity of threshold $\beta$ in Eq. (\ref{eq:vadthr}) of rVAD for voice activity detection.
Table \ref{table:rats_res_sensitivity} presents the results on the RTAS database when using different values of $\beta$; other than this, the default configuration of rVAD is used.
It is observed that increasing the value of the VAD threshold $\beta$ makes rVAD more aggressive, i.e. increased $P_{miss}$ and decreased $P_{fa}$, and thus less frames are classified as speech. The $\beta$ value should be chosen according to the application in hand. Furthermore, the results show that rVAD performs well in a wide range of values of threshold $\beta$, demonstrating its advantage of stability. 

\begin{table}[H]
\caption{\it Performance of rVAD with various threshold $\beta$ values on the RATS database.}
\begin{center}
\begin{tabular}{|lllcl|}\cline{1-5}
 Method &               Threshold & $P_{miss}(\%)$ & $P_{fa}(\%)$ & FER \\ 
                               & ($\beta$)  &        &          & (\%)    \\  \hline
                               & 0.1        & 8.85 & 10.58   & 9.86      \\
                               & 0.2        & 11.36 & 8.16   & 9.48      \\
                               & 0.3        & 14.04 & 6.60   & 9.68   \\ 
rVAD       & 0.4(default)& 17.00& 5.52   & 10.27   \\
(MSNE)                         & 0.5        & 20.33 & 4.70   & 11.17   \\ 
                               & 0.6        & 24.16 & 4.02   & 12.36   \\
                               & 0.7        & 28.79 & 3.44   & 13.94  \\ \hline                                              
\end{tabular}
\end{center}
\label{table:rats_res_sensitivity}
\end{table}	                    

\subsection{VAD results for NIST 2015 OpenSAD challenge}

In \cite{Kinnunen+2016}, the rVAD method is compared with several methods on the NIST 2015 OpenSAD challenge \cite{NISTSAD2015} that is also based on the RATS database. Table \ref{table:Happy_result} cites results for several methods from \cite{Kinnunen+2016}, our joint work with several other teams. The GMM-MFCC and GMM-PNCC methods are similar to \cite{VQVAD}, but using GMMs trained with maximum-likelihood instead of codebooks, which leads to some improvement. PNCCs \cite{PNCC2012} are known to be robust against noise. The i-vector based VAD is a supervised VAD \cite{Khoury+2016}.  rVAD shows a performance substantially better than those of Sohn \emph{et al.}, GMM-MFCC and GMM-PNCC, and a performance very close to that of the supervised i-vector method trained on the NIST 2015 OpenSAD challenge.  

\begin{table}[H]
\caption{\it VAD results \cite{Kinnunen+2016} of several methods on Dev data of NIST 2015 OpenSAD challenge. }
\begin{center}
\begin{tabular}{|l|l|l|l|}\cline{1-4}
 Methods                          & $P_{fa}(\%)$   & $P_{miss}(\%)$ & DCF \\ \hline
GMM-MFCC \cite{Kinnunen+2016}                                 &  6.15    &  43.17  & 0.1540    \\ 
GMM-PNCC  \cite{Kinnunen+2016}                     & 7.72     &  17.14  & 0.1008 \\
Sohn et al. \cite{Sohn1999a}    & 6.35     &  38.89  & 0.1449       \\
i-vector (supervised) \cite{Khoury+2016}             & 2.77     & 10.09   & {\bf 0.0460}     \\
rVAD (MSNE-mod)                         & 4.78     & 5.75   & 0.0502      \\ \hline  
  \end{tabular}
 \end{center}
 
\label{table:Happy_result}
\end{table}

\subsection{Speaker verification results on the RATS database}
In \cite{PlchotMMDMCGHMMSSTTZZ13}, the rVAD method was applied as a preprocessing step of text-independent speaker verification (TI-SV) systems built for the RATS database under the DARPA RATS project. The TI-SV systems use $60$-dimension MFCCs as features and  $600$-dimension i-vector \cite{Deka_ieee2011} together with probability linear discriminate analysis (PLDA) based scoring as the speaker verification back-end. Praat pitch extration \cite{Boersma2009} was used for the rVAD method in this experiment. rVAD was first compared with a neural network based supervised VAD method developed by Brno University of Technology (BUT-VAD) \cite{Ng2012a}. The neural network deployed in BUT-VAD has $9$ outputs for speech and  $9$ outputs non-speech, each of which corresponds to one of the $9$ channels (one source and $8$ retransmitted). The outputs are smoothed and merged into speech and non-speech segments using a hidden Markov model (HMM) with Viterbi decoding. The neural network was trained on RATS data defined for the VAD task. Another supervised VAD method is GMM-PLP-RASTA where perceptual linear predictive (PLP) coefficients with RASTA-based \cite{Hermanksy94} cepstral mean normalization being applied are used as the feature and two 2048-component GMMs (one for speech and another for non-speech), trained on RATS, are used as the models for VAD. 

  Evaluation was conducted on a development set of $30$s-$30$s enrolment and test condition and the evaluation criterion for speaker verification is equal error rate (EER) \cite{PlchotMMDMCGHMMSSTTZZ13}. 
 The unsupervised rVAD (MSNE-mod) method, the supervised BUT-VAD and GMM-PLP-RASTA methods\cite{Ng2012a} yield EERs of $5.6\%$, $5.4\%$, $6.7\%$, respectively.
  rVAD (MSNE-mod) performs marginally worse than the supervised BUT-VAD trained on the RATS database, but better than the supervised GMM-PLP-RASTA also trained on the RATS database. For more details about the systems see \cite{PlchotMMDMCGHMMSSTTZZ13}.

\section{Experiments on the RedDots  database for TD-SV}
\label{sec:RedDots}
In this section, we compare rVAD and rVAD-fast with other VAD methods in the context of  text-dependent speaker verification (TD-SV) on the male part-01 task of the RedDots $2016$ challenge database \cite{RedDots}, which consists of short utterances and is one of most used databases for speaker verification. The database was collected in many countries and mostly in office environments, hence introducing great diversity in terms of speakers and channels. There are $320$ target models for training and each target has $3$ speech signals/sessions for building their particular model. For the assessment of TD-SV, there are three types of non-target trials: target-wrong (TW) ($29,178$), impostor-correct (IC) ($120,086$) and impostor-wrong (IW) ($1,080,774$).  

The modelling method for SV adopts Gaussian mixture model-universal background model (GMM-UBM) since GMM based methods are known to outperform the i-vector technique for  SV with short utterances. MFCCs of $57$ dimensions  (including static, $\Delta$ and $\Delta\Delta$) with RASTA filtering \cite{Hermanksy94} are extracted from speech signals with a $25$ ms hamming window and a $10$ ms frame shift. After VAD, detected speech frames are normalized to have zero mean and unit variance at utterance level. A gender independent GMM-UBM, consisting of $512$ mixtures and having diagonal covariance matrices, is trained using non-target data from TIMIT over 630 speakers (6300 utterances). Target models are derived from the GMM-UBM with $3$ iterations of MAP adaptation (with relevance factor $10.0$ and only applied to Gaussian mean vectors) using the training data of the particular target model. During the test, feature vectors of a test utterance is scored against the target model and GMM-UBM to calculate log likelihood ratio.  
Table \ref{table:table_td} presents the TD-SV performance for different methods which are generally proposed for the SV. 

\begin{table}[H]
\caption{\it Performance of VAD methods for text-dependent speaker verification on RedDots part-01 (male)}
\begin{center}
\begin{tabular}{|l|ccc|c|}\cline{1-5}
Methods     & \multicolumn{3}{c|}{[\%EER/minDCFx100]}  & Average \\
            &Target-wrong  & Impostors-correct & Impostors-wrong & (\%EER/minDCF)\\
            & (TW)         & (IC)              & (IW)            &            \\ \hline 
 no VAD                 & 6.36/3.071 &2.82/1.400 & 1.26/0.525 & 3.48/1.665  \\            
 Kaldi Energy VAD \cite{Povey_ASRU2011}       & 6.26/2.662 &3.62/1.625 &1.67/0.618  & 3.85/1.635 \\ 
 Sohn et al. \cite{Sohn1999a}            & 4.78/2.492 &{\bf 2.40}/{\bf 1.200} &1.01/0.295  & 2.73/1.329 \\ 
 VQVAD \cite{VQVAD}                  & {\bf3.70}/1.652  & 2.94/1.520      & 0.89/0.284 & {\bf 2.51}/1.152 \\ 
                     &            &           &            &            \\
 \emph{rVAD(MSNE, default) }            & 3.79/1.572 &2.93/1.328 &1.01/0.273  & 2.58/1.058 \\
 rVAD(MMSE)             & 4.16/1.523 &3.02/1.329 &{\bf 0.92}/0.300  &2.70/1.050 \\
 rVAD(MSNE-mod)       & 3.82/{\bf 1.498} &2.83/1.290 &0.98/0.284  & 2.54/{\bf 1.024}  \\
                        &            &           &            &            \\
 rVAD-fast(MSNE)        & 4.10/1.682 &2.86/1.228 &{\bf 0.92}/{\bf 0.272}  &2.63/1.061 \\
 rVAD-fast(MMSE)        & 5.09/2.242 &2.74/1.276 &1.14/0.383  & 2.99/1.300 \\
 rVAD-fast(MSNE-mod)   &3.97/1.705	 & 2.64/1.225&0.95/0.283  & 2.52/1.071 \\ \hline
 \end{tabular}
   \end{center}
  \label{table:table_td}
\end{table}
 
 The experiment results in Table \ref{table:table_td} show that that all VAD methods (except for Kaldi VAD) are able to outperform the system without VAD  either in terms of EER or minDCF. This observation is in line with the well known fact that VAD is useful in speaker verification. rVAD overall performs better than Sohn \emph{et al.}, Kaldi Energy VAD and is comparable to or marginally better than VQVAD (almost identical in EER and slightly better in minDCF). rVAD-fast shows a similar performance to rVAD in terms of SV, although its VAD performance (as observed in Table \ref{table:tableSFT}) is worse than that of rVAD. Considering the huge VAD performance gap (by a factor of more than three in EER) between rVAD and VQVAD as shown in Tables \ref{table:table1} and \ref{table:rats_res_200}, we conclude that superior performance in VAD does not necessarily translate into an improvement in SV performance. This could be explained by the fact that SV is a task of making one single decision based on the entire sequence/utterance, which differs from the VAD task where each short segment matters. VQVAD is specially optimized for SV. 

\subsection{Sensitivity of rVAD threshold $\beta$ on speaker verification performance}
Table \ref{table:table_td_sensitivity} shows the effect of varying threshold value $\beta$ in Eq. (\ref{eq:vadthr}) of rVAD on TD-SV performance. 
MSNE is used for the second pass denoising and both passes are applied. The results show that the performance of rVAD does not change rapidly with changing the value of $\beta$, demonstrating the stability of rVAD. The results, which are obtained but not included in Table \ref{table:table_td_sensitivity}, also show that rVAD with MSNE-mod performs slightly better than rVAD with MSNE.

\begin{table}[H]
\caption{\it Performance of rVAD with various threshold $\beta$ values for speaker verification on RedDots part-01 (male).}
\begin{center}
\begin{tabular}{|l|c|ccc|c|}\cline{1-6}
Methods     & Threshold & \multicolumn{3}{c|}{[\%EER/minDCFx100]}  & Average \\
            & ($\beta$) &Target-wrong  & Impostors-correct & Impostors-wrong & (\%EER/minDCF)\\
            &           & (TW)         & (IC)              & (IW)            &            \\ \hline 
            & 0.1       & 4.08/1.591   & 2.77/1.243        & 1.07/0.266      &  2.64/1.033 \\
            & 0.2       & 3.91/1.534   & 2.80/1.242        & 1.11/0.278      & 2.61/{\bf1.018}  \\
            & 0.3       & 3.86/1.547   & 2.82/1.249        & 1.01/0.275      & {\bf2.56}/1.024   \\
 rVAD       & 0.4 (default)& 3.79/1.572   & 2.93/1.328        & 1.01/0.273      & 2.58/1.058 \\
 (MSNE)     & 0.5       & 4.36/1.646   & 3.26/1.451        & 1.14/0.331      & 2.92/1.143    \\
            & 0.6       & 4.19/1.674   & 3.20/1.460        & 1.26/0.325      & 2.88/1.153 \\
            & 0.7       & 4.45/1.735   & 3.54/1.540        & 1.29/0.388      & 3.09/1.221 \\ \hline 
  \end{tabular}
  \end{center}
  \label{table:table_td_sensitivity}
\end{table}

\subsection{Text-dependent speaker verification under noise conditions}
\begin{figure}[H] 
\centering{\includegraphics[width=14.2cm,height=7.5cm]{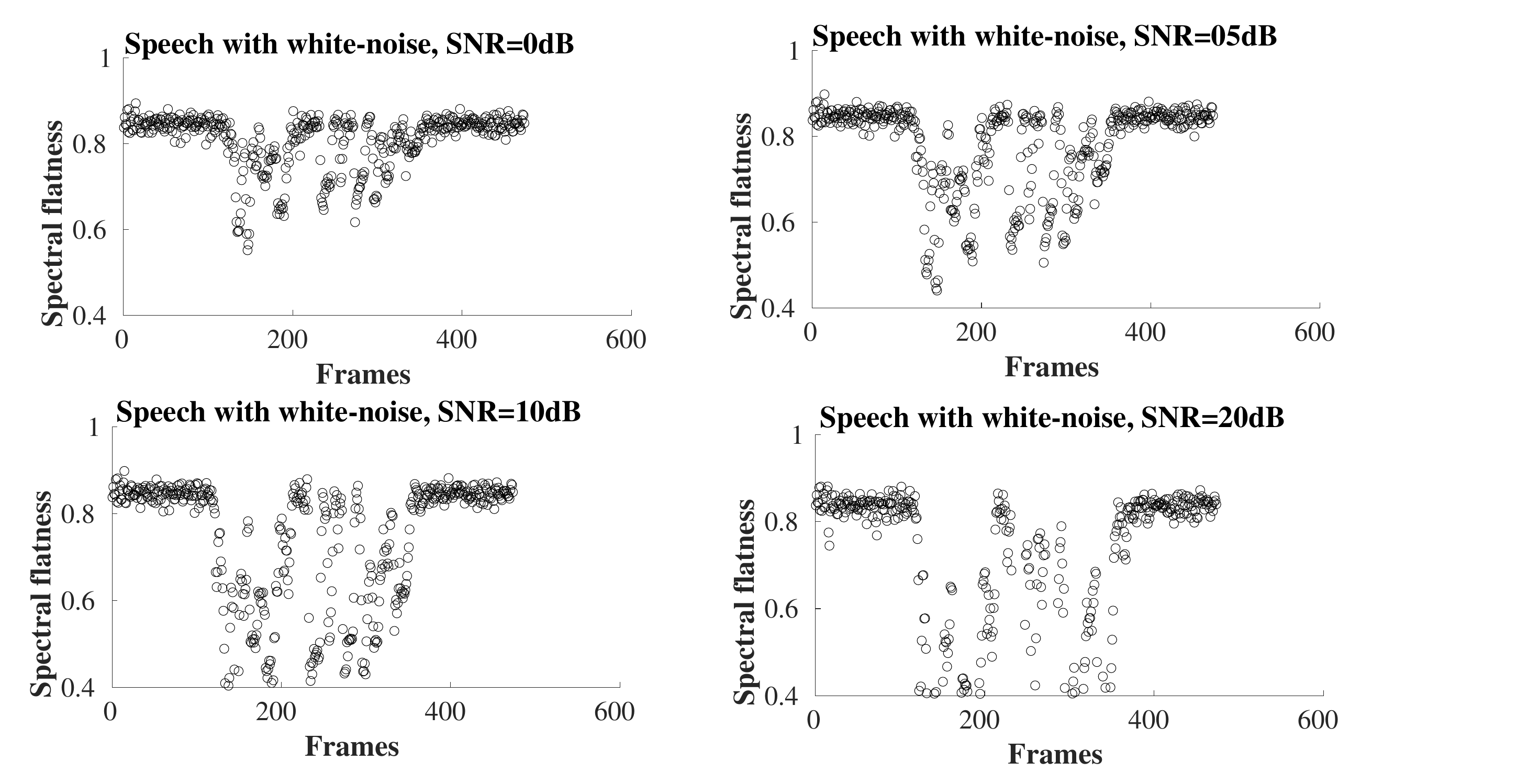}}
\caption{\it Scatter plots of SFT values of a speech signal with white-noise for different SNR values.}
\label{fig:ft_whiteNoise}
\end{figure}

We further evaluate the performance of VAD methods for TD-SV under mismatched conditions, where noisy test utterances are scored against the speaker models trained under clean condition (office environment). In order to cover different real-world scenarios,  various types of noise are artificially added to the test data with various SNR values. The scaling factor is calculated using the ITU speech voltmeter \cite{G191}. TD-SV results of different VAD methods are presented in Table \ref{table:table_noisy}. It is observed that TD-SV performances are significantly degraded under noisy conditions as also expected. rVAD achieves mostly lower EER values than those of Kaldi and Sohn et al. and \emph{No-VAD} (i.e., without using VAD) over different noise types and SNR values, but performs slightly worse than VQVAD that is specially designed for SV. 
Sohn \emph{et al.} VAD provides decent improvement as well. Kaldi Energy VAD degrades the performance compared with No-VAD, as in the case of clean condition. 

rVAD-fast gives comparable performance to that of rVAD under noise type of babble, market and car, but it does not work under white noise as the spectral flatness measure is severely affected by white noise. 
A close analysis shows that for  white noise from $0 dB$ through $10dB$, spectral flatness values are mostly close to $1.0$, as illustrated in Fig.\ref{fig:ft_whiteNoise}, due to the similar amount of power in all spectral bands. Therefore, the threshold value of $\theta_{sft}=0.5$ does not output any speech frames with pitch (and thus no speech frames) for most of noisy test trails in SV and in this experiment, we consider these trails without speech frames as an SV error or misclassification. Numbers of these trails (without any speech frame being detected) are shown in parenthesis in Table \ref{table:table_noisy}. When calculating EER, genuine/true trials without speech frames are directly rejected (false rejection) by assigning the lowest score value available in the non-target trials to these genuine trials   and vice-versa, namely non-target trails without speech frames are directly accepted (false alarm) by assigning the highest score value available in the genuine trials to these non-target trials.

\begin{table}[h]
\caption{\it TD-SV performance of different VAD methods under noisy test environments on RedDots part-01 (male). Numbers in parenthesis show numbers of test utterances (out of total $3854$ unique utterances) that do not yield any speech frames by the respective VAD methods. }
\scriptsize{
\begin{center}
\begin{tabular}{|l|l|llll|lll|lll|}\cline{1-12}
Noise     & SNR & \multicolumn{10}{c|}{ \% Average EER [TW, IC, IW] across VAD methods}  \\ \cline{3-12}
type & (dB) & no & Kaldi & Sohn & VQ        &\multicolumn{3}{|c|}{rVAD}& \multicolumn{3}{c|}{rVAD-fast} \\ \cline{7-12}
         &    & VAD &     & et al. & VAD    & MSNE  & MMSE  & mod.      & MSNE    & MMSE       & mod.   \\ 
         &    &     &     &       &                          & (default)  &  & MSNE      &         &            & MSNE  \\  \hline
 Clean   & -  & 3.48& 3.85& 2.73 & {\bf 2.51}                          &2.58   & 2.70  & {\bf 2.54} & 2.63(2)& 2.99(2)    & {\bf 2.52}(2)  \\  \hline  
 White  & 0   & 35.23& 37.82 & 37.05(119)&30.18              & 34.46      & 34.44 & 35.29(65)      & 99.34(3796) & 99.38(3796) & 99.29(3796) \\
        & 05  & 26.78& 30.63&26.68(33)&  21.98                &25.49      & 24.82 & 25.58(11)     & 68.05(2086) & 68.05(2080) & 67.84(2079) \\
        & 10 & 18.88 & 22.19&17.08 & 14.57                    &16.42  & 16.90      & 17.22      & 25.26(333)  & 25.26(326)  & 25.25(335) \\ 
        & 15 & 12.47 & 15.28& 10.50 &  9.28                    &10.20 & 10.33      & 10.41      & 13.04(61)   & 13.13(81)   & 13.30(63)  \\
        & 20 & 8.25  & 9.90 & 6.59  & 5.97                   &6.64       & 6.60  & 6.62       & 6.85(3)    & 7.12(3)    & 7.05(3) \\ 
               \hline 
Babble  & 0  & 35.02    & 34.42  & 33.25 & 30.91               &32.88      & 32.78 & 33.10      & 33.01   & 33.44   & 33.21 \\
        & 05  & 25.17   & 24.19  & 22.89 & 20.78              & 21.84 & 22.32      & 22.36      & 22.84   & 23.79   & 21.90  \\
        & 10 & 16.45    & 15.61  & 14.17 & 12.48              &13.49 & 13.72      & 13.60      & 14.01(1)& 14.43   & 13.61 \\ 
        & 15 & 10.30    & 10.26  & 8.48  & 7.55              &8.19  & 8.47       & 8.24       & 8.37(2) & 8.65    & 8.41  \\
        & 20 & 6.34     &  6.94  & 5.24  & 4.66              &5.28   & 5.31           & 5.17  & 5.46(2) & 5.43(2)    & 5.59(2)  \\ \hline
Market  & 0  & 25.39    & 26.24  & 24.67 & 22.79               &23.83  & 24.15      & 24.09(1)   & 24.37   & 24.49   & 23.94  \\
        & 05  & 16.84   & 17.41  & 15.51 & 14.09               &15.07  & 15.33      & 15.16      & 15.81   & 15.38   & 15.11  \\
        & 10 & 10.58    & 11.11  & 9.02  & 8.40               &8.84  & 9.06       & 9.11       & 9.23    & 9.37    &  9.33   \\ 
        & 15 & 6.58     & 7.45   & 5.65  & 5.06               &5.61   & 5.90       & 5.73       & 6.00(1) & 6.06(1)    & 5.97(1)  \\ 
        & 20 & 4.82     & 5.44   & 3.90 & 3.60               &3.99   & 4.10       & 3.98       & 4.03(2) & 4.21(2) & 4.04(2)  \\\hline
 Car    & 0  &  4.10    & 5.60   & 3.70 &  3.53                &3.74   & 3.95       & 3.75       & 3.75        & 3.74    & 3.69 \\ 
        & 05  & 3.67    & 5.01   & 3.20 & 3.09                &3.23 & 3.39       & 3.24       &  3.24       & 3.23    & 3.18 \\
        & 10  & 3.33   & 4.61   & 3.00  & 2.90                 &2.90& 3.05 & 2.96             & 2.92        & 2.93    & 2.88 \\
        & 15  & 3.24   & 4.36   & 2.80  & 2.72               &2.77 & 2.85  &   2.71         &  2.77       & 2.78    & 2.71 \\
        & 20  & 3.18   & 4.18   & 2.70  & 2.58                &2.62 & 2.71  & 2.60        & 2.63(1)        & 2.70(1)  & {\bf2.56}(1) \\   \hline
Average &     & 13.83 &	14.93  & 12.80  & 11.35                &12.37&12.50 &12.54 &18.54 &18.67 &18.44     \\ \hline                
\end{tabular}
\end{center}
}
\label{table:table_noisy}
\end{table}

 \subsection{Sensitivity of rVAD threshold $\beta$ on speaker verification performance under noisy conditions}  
 Table \ref{table:table_noisy_sen} shows the effect of varying threshold $\beta$ in Eq. (\ref{eq:vadthr}) of rVAD on the performance of TD-SV under noisy test conditions. 
 It is observed that rVAD is stable towards varying the threshold value.
 
 \begin{table}[H]
 \caption{\it Performance of rVAD with various threshold $\beta$ values for speaker verification on RedDots part-01 (male) under noisy conditions.}
 \footnotesize{
 \begin{center}
 \begin{tabular}{|l|l|ccccccccc|}\cline{1-11}
 Test          & SNR & \multicolumn{9}{c|}{ Threshold ($\beta$) [\% average EER (TW, IC, IW)] }  \\ 
condition    & (dB) & 0.1 & 0.2 & 0.3 & 0.4 (default) & 0.5 & 0.6 & 0.7 & 0.8  & 0.9  \\ \hline
 White       & 0    & 34.55&34.37&34.16&34.46&34.41&33.91&{\bf33.47}    & 34.27 & 34.76  \\
             & 05   & 25.79&25.26&25.09&25.49&25.40&25.28&{\bf24.47}    & 25.62 & 25.39 \\
             & 10   & 17.40&16.87&16.71&16.42&16.75&16.95&{\bf16.11}    & 17.27 & 17.63\\
             & 15   & 10.87&10.66&10.61&10.20&10.42&10.65&{\bf10.10}    &11.20  & 11.69\\
            & 20   & 6.79  &6.71& {\bf6.49} &6.64 &6.79 &6.74 &6.64     &7.45   & 7.78\\ \hline 
 Babble     & 0   & 33.79 &33.02&33.07 &32.88&32.81&32.34&{\bf31.71}    & 31.63 & 32.13\\
            & 05  &23.85  &23.45&22.65 &21.84&21.86&21.42&{\bf20.84}    & 21.49 & 21.95 \\
            & 10  &14.82  &14.42&13.88 &13.49&13.66&13.15&{\bf12.79}    & 13.60 & 13.72\\
            & 15  &8.92   &8.56 &8.36  &8.19 &8.30 &7.97 & {\bf7.95}    & 8.37  & 8.59\\
            & 20  &5.82   &5.54 &{\bf5.26}  &5.28 &5.28 &5.30 & 5.37    & 5.86  & 6.28\\ \hline 
 Market     & 0  &24.43  &23.87 &23.96 &23.86&23.97&{\bf23.08}& 23.36   & 24.09 & 24.53\\
            & 05 &15.54  &15.22 &14.75 &15.07&15.18&14.68& {\bf14.36}   & 15.21 & 15.55\\
            & 10 &9.62   &9.02  &{\bf8.76}  &8.84&9.02  &9.01 &8.83     & 9.71  & 9.88\\
            & 15 &6.21   &5.85  &5.71  &{\bf5.61} &6.02 &5.86 &5.94     & 6.61  & 6.79\\
            & 20 &4.20   &4.10  &{\bf3.99}  &{\bf3.99} &4.25 &4.15 &4.51 & 4.82 & 5.09   \\ \hline 
 Car        & 0  &3.74   &{\bf3.69} &3.76   &3.74 &4.05 & 4.05&4.30      & 5.02 & 5.21\\
            & 05 &3.26   &{\bf3.21} &3.26   &3.23 &3.60 &3.60 &3.78      & 4.51 & 4.86\\
            & 10 &2.94   &{\bf2.87} &2.97   &2.90 &3.30 &3.30 &3.52      & 4.17 & 4.49\\
            & 15 &{\bf2.71}   &2.75 &2.72   &2.77 &3.07 &3.09 & 3.29     & 3.97 & 4.14\\
            & 20 &2.65   &2.63 &2.63   &{\bf2.62} &2.92 &2.92 & 3.14     & 3.79 & 4.00\\ \hline 
Average     &    &12.89 &12.60 &	12.43 &	12.37 &	12.55 &	12.37&	{\bf12.22}&	12.93 & 13.22   \\ \hline  \end{tabular}
\end{center}
}
\label{table:table_noisy_sen}
\end{table}

\section{Conclusion}
\label{sec:conclusion}
In this paper, we presented an unsupervised segment-based robust voice activity detection (rVAD) method for voice activity detection (VAD). It consists of two-pass denoising, extended pitch segment detection, and voice activity detection. The first pass denoising uses pitch as a speech indicator to remove high-energy noise segments detected by using \emph{a posteriori} signal-to-noise ratio (SNR) weighted energy difference, while the second pass denoising attempts to remove more stationary noise using speech enhancement methods. Then, extended pitch segments are found. In the end, \emph{a posteriori} SNR weighted energy difference is applied to extended pitch segments of the denoised speech signal for VAD. We evaluated the performance of the proposed rVAD method for VAD and speaker verification (SV) tasks on several diverse databases containing a large variety of noise conditions and compared rVAD against 16 VAD methods including both supervised and unsupervised methods. Experiment results show that the proposed method is compared favourably with a number of existing methods. It is worth to emphasize that rVAD obtained the promising performances across databases and tasks by using the same parameters, indicating the good generalization ability of rVAD. It can be concluded that pitch is a good indicator or anchor for locating speech segments and \emph{a posteriori} signal-to-noise ratio (SNR) weighted energy difference is an effective measure for segmenting speech in noisy environment. Furthermore, a VAD targeted towards SV is able to perform well even though its VAD performance does not.

In addition, we presented a modified version of the rVAD method, called rVAD-fast, where computationally time-consuming  pitch extraction is replaced by computationally efficient spectral flatness calculation. The modified version significantly reduces the computational complexity at the cost of moderate inferior VAD performance, which is an advantage when processing a large amount of data and running on resource-limited devices. rVAD-fast, however, breaks down under white noise and therefore it should be used with caution. It has shown to work well under babble, market and car noise and under clean condition. One further finding is that spectral flatness is a good indicator for whether or not there is pitch in a segment as long as the signal is not severely corrupted by white noise.  

Overall, it can be concluded that rVAD performs well in both clean and noisy conditions and for both VAD itself and SV. The generalization ability across databases, noisy conditions and tasks was proofed as well. 

Future work includes investigating the optimal configurations of rVAD for different applications. The performance of rVAD on automatic speech recognition is worth to study as well.

\section{Acknowledgement}
Zheng-Hua Tan wishes to thank Dr. James Glass for hosting him in 2012 and 2017 at MIT where the work was done in part. 

This work is partly supported by the iSocioBot project, funded by the Danish Council for Independent Research - Technology and Production Sciences (\#1335-00162), and the Horizon 2020 OCTAVE Project (\#647850), funded by the Research European Agency (REA) of the European Commission.

\bibliographystyle{elsarticle-num}
\bibliography{Reference}

\end{document}